# Optical imaging of antiferromagnetic domains in ultrathin CoO(001) films

Jia Xu[1], Haoran Chen[1], Chao Zhou[1], Dong Shi[1], Gong Chen[2], Yizheng Wu[1*]

[1]Department of Physics and State Key Laboratory of Surface Physics, Fudan University, Shanghai 200433, China

[2]Department of Physics, University of California, Davis, California 95616, USA

**Abstract**

Antiferromagnetic (AFM) domains in ultrathin CoO(001) films are imaged by a wide-field optical microscopy using magneto-optical birefringence effect. The magnetic origin of observed optical contrast is confirmed by the spin orientation manipulation through exchange coupling in Fe/CoO(001) bilayer. The finite size effect of ordering temperature for ultrathin single crystal CoO film is revealed by the thickness and temperature dependent measurement of birefringence contrast. The magneto-optical birefringence effect is found to strongly depend on the photon energy of incident light, and a surprising large polarization rotation angle up to 168.5 mdeg is obtained from a 4.6 nm CoO film with a blue light source, making it possible to further investigate the evolution of AFM domains in AFM ultrathin film under external field.

I. **Introduction**

Antiferromagnet is attracting considerable attention due to its potential in future spintronic applications [1-7]. Antiferromagnet has been applied as the pinning layer in spintronics devices for decades, and recently was considered to replace the ferromagnet as the active spin-dependent information element in the next generation of spintronic devices, due to the robustness against perturbation from external magnetic fields, the absence of stray fields, ultrafast dynamics, and considerable magneto-transport effects in antiferromagnetic (AFM) materials with wide variety [8]. Current-induced switching of AFM spins in both metallic [2-4] and insulating [5-7] AFM systems has been recently reported, making it possible to store information in AFM spintronics devices. Most experimental investigations on the current-induced switching of AFM domains were utilized by the anisotropic magnetoresistance, spin Hall magnetoresistance, and the related planar Hall resistance, but the electronic signals induced by the current pulse are not necessary correlated with the evolution of AFM domain states [9-11]. So, there is an urgent need to directly measure the AFM domain in real space during the operation of electric current for further understanding the mechanism of current-induced AFM domain switching.

To date, the most common technique to study the AFM domains is the photoemission electron microscopy (PEEM) based on X-ray magnetic linear dichroism effect (XMLD) [3,5,12] However, the XMLD-PEEM measurement requires the access to large synchrotron facilities, and is also difficult to be incorporated simultaneously with the external electric current and magnetic fields. Recently, we reported that the magneto-optical birefringence effect can be applied to image the AFM domains in NiO thin films by a tabletop Kerr microscope [13], which is more accessible than XMLD-PEEM and can work under external fields. Thus, it would be fundamentally interesting to explore whether this imaging technique can be applied to the other AFM materials.

CoO is another important AFM material with a bulk Néel temperature ($T_N$) of 293 K [14-16]. Although the AFM spins in bulk CoO align along the <117> directions, the CoO films show different spin structures due to the strain effect from the substrate [17,18]. Single crystalline CoO film can be well epitaxied on MgO(001) substrates [17,19], with the spins aligning in the film plane due to the tensile stress in the CoO film [19]. The CoO AFM spins in Fe/CoO(001) bilayers has been measured by XMLD-PEEM, where the Fe FM spins and CoO AFM spins are perpendicularly coupled [20,21]. The AFM domain nucleation and propagation in Fe/CoO(001) under magnetic field has been interpreted indirectly through the measurement of ferromagnetic (FM) properties of Fe layer utilizing the magneto-optical Kerr effect [19]. Thus, to further understand the exchange coupling between CoO and FM layer, a direct imaging of CoO AFM domain in the presence of an external field can provide pivotal information. Recently, utilizing a femtosecond pump–probe method, Zhen et al. demonstrate the Néel vector dynamics in CoO film due to large magneto-optical Voigt effect [22]. The optical birefringence effect has been applied to image the AFM domains in bulk CoO crystal by light transmissions [15,17,23], thus it is also feasible that the AFM domain structure of ultrathin CoO film can be studied by the optical birefringence effect in the reflection geometry.

In this work, we report the studies on the AFM domains in single crystal CoO thin films grown on MgO(001) substrates with the magneto-optical birefringence effect. By manipulating the AFM Néel vector through the exchange coupling in Fe/CoO bilayer, we confirm the AFM order based origin of the observed optical contrast. Through the systematical studies on the temperature and thickness dependence of the magnetic contrast, we prove that the AFM domains can be observed down to 1.5 nm, and the thickness dependent $T_N$ of CoO film follows the finite size effect. Finally, we discover that the magneto-optical birefringence effect has strong dependence on the wavelength of the incident light, and the large polarization rotation angle up to 168.5 mdeg is quantified for a 4.6 nm CoO film. Our studies demonstrate that the magneto-optical birefringence effect is an effective tool to study the spin properties in AFM ultrathin films under external field, which could be helpful for the development of AFM spintronics.

**II. Experimental details**

The single-crystalline CoO film were grown on MgO (001) substrates in an ultrahigh vacuum system by molecular beam epitaxy (MBE) [19,24]. The MgO(001) single-crystal substrates were annealed at 600 ℃ for half an hour inside an ultrahigh vacuum chamber. A 6 nm-thick MgO seed layer was deposited at 500 ℃ before the CoO growth. The CoO film was then grown by evaporating Co under an oxygen pressure of $2.0 \times 10^{-7}$ Torr at room temperature [19,24]. For thickness dependent measurements, the CoO film is grown with a wedged shape by moving the substrate behind a knife-edge shutter. All the samples were capped with a 5 nm MgO as a protective layer before taken out of the vacuum chamber. The film thickness was determined by the deposition rate, which was monitored with a calibrated quartz thickness monitor. Sharp reflection high energy electron diffraction patterns reveal excellent epitaxy growth of CoO film with the lattice relation of Fe[110](001)//CoO[100](001)//MgO[100](001) [19,24].

The CoO AFM domains are imaged with a commercial Evico magneto-optic Kerr microscope [25, 26]. A white-color light-emitting diode (LED) with a wide wavelength range between 420 nm and 650 nm is mostly used as the light source in this study. During the measurements, the sample temperature can be varied between 77 K and 330 K. The magnetic field up to 1000 Oe is applied by a rotatable electromagnet. Our Kerr microscope is also equipped with a red LED source with the wavelength of ~650 nm and a blue LED source with the wavelength of ~455 nm, thus we also measured the optical birefringence contrasts with different LED sources.

**III. Results and discussion:**

Figure 1(a) presents the measurement geometry for imaging the CoO AFM domains utilizing the magneto-optical birefringence effect, and the detailed procedure on the AFM domain measurements can be found in Ref. 13. It is well known that the CoO AFM spins lie in the film plane with the easy axis along <110> direction, which has been determined by XMLD [19,27-29]. Due to the in-plane four-fold lattice symmetry of CoO(001) film, the in-plane Néel vectors in different CoO AFM domains align along either [110] or [1$\bar{1}$0] directions. A linear-polarized light is normally incident on the sample with the light polarization 45° away from the <110> direction of the CoO. As demonstrated in Ref. [13], the polarization of the reflected light from an AFM domain will rotate by a certain angle $\theta_v$ due to the birefringence effect, and two orthogonal AFM domains are expected to induce opposite the rotation of the polarization. In a Kerr microscope, the reflected light passes through an analyzer before being detected by the CCD camera. Usually the analyzer is set with a small offset angle $\theta_P$ from the extinction position, after passing through the analyzer, the light intensities (I) from the two orthogonal AFM domains will be proportional to $\sin^2(\theta_P - \theta_v)$ and $\sin^2(\theta_P + \theta_v)$ respectively. Such an intensity difference gives rise to the optical contrasts, which allows distinguishing the 90° AFM domains in CoO films.

Figures 1(b) and (c) represent the typical optical images from a 5.5 nm CoO film with $\theta_P = -7°$ and $\theta_P = +7°$ measured at 77 K, which show the opposite contrast. The contrast due to the optical birefringence effect can be better identified by calculating the asymmetry, i.e. $I_{\text{asy}} = \frac{I(+\theta_p) - I(-\theta_p)}{I(+\theta_p) + I(-\theta_p)}$, as shown in Fig.1(d), and clear black and white domains can be observed [13]. The contrast can be attributed to the AFM domains with the AFM magnetization aligned along <110> and <1$\bar{1}$0> axis due to the in-plane four-fold symmetry of CoO(001).

Next, we further confirm that the measured optical contrast of CoO films in Fig. 1 originates from the AFM domains. One common way is to directly compare the optical image with the XMLD-PEEM image [13], which requires the XMLD-PEEM beamtime in the synchrotron facilities. It is well known that CoO has the G-type AFM spin structure with a compensated (001) surface, and in an Fe/CoO(001) system the Fe FM spins and CoO AFM spins are perpendicularly coupled [19], thus the CoO spin orientation in Fe/CoO(001) can be manipulated by field cooling. Next, we prepare a 5 nm thick CoO film, and half of it is covered with a 2 nm Fe film for comparison, see sample structure in Fig. 2(a). We measure the optical birefringence effect near the boundary between pure CoO film and Fe/CoO bilayer after the field cooling. The field cooling is performed from 330 K down to 77 K within a field of 1000 Oe along different directions, then the birefringence images are measured at zero field, as shown in Figs. 2(b)-(d). For $H_{FC} \parallel [110]$ in Fig. 2(b), there is only uniform dark contrast in the Fe/CoO region, but the multi-domains can be observed in the pure CoO region. For $H_{FC} \parallel [\bar{1}10]$ in Fig. 2(c), the domain patterns in the pure CoO region remain unchanged, but the contrast in Fe/CoO region is reversed. It is expected that the alternating demagnetization field may generate multi-domains in the Fe film. Then we apply the AC demagnetization field along <100> at 330K, then cool down the sample down to 77 K at zero field. We find that the Fe/CoO region shows the dominating black contrast, but has some small areas with white contrast, as shown in Fig. 2(d). The applied AC field may not be perfectly aligned along <100>, which can result in the large Fe FM domain and CoO AFM domain. Our results in Fig. 2 can conclude that the change of the birefringence contrast is mainly attributed to the orthogonal AFM spins rotated by the exchange coupling from the Fe film during the field cooling process. Since the CoO Néel vector is perpendicular to the Fe magnetization in Fe/CoO(001) bilayer [19], the white birefringence contrast refers the CoO Néel vector along <110> axis, and the CoO Néel vector along <1$\bar{1}$0> axis gives the black contrast, thus the AFM spin orientation in different AFM domains can be determined by the birefringence effect. We note that the birefringence contrast in Fe/CoO bilayer mainly comes from the CoO layer, since our measurement shows that the birefringence effect from a 5 nm Fe film is negligibly small [30-32].

We then investigate the thickness-dependence of the CoO birefringence images at 77 K and 290 K respectively. The measurement is performed on a CoO-wedge sample with a thickness range of 0-12 nm and a slope of 2 nm/mm. Figures 3(a)-(d) show the domain images with different CoO thicknesses at 290 K. When the film is thinner than 1.5 nm, the image contrast is too weak to be convincingly distinguished, indicating that the $T_N$ of ~1.5 nm CoO film is below 290 K. For the film thicker than 1.5 nm, the AFM order gradually builds up, and the contrast increases with film thickness. Figures 3(e)-(h) show the domain images measured at 77 K at the same areas, with the domain contrast much stronger than that at 290 K. The clear domain contrast of the 1.5 nm thick CoO also indicates that its ordering temperature $T_N$ is higher than 77 K.

We further quantify the thickness dependence of the contrast at 77 K and 290 K, as shown in Fig. 4(a). The domain contrast is almost linearly dependent on the CoO thickness. The optical birefringence effect in NiO/MgO(001) [13] also has the similar linear thickness dependence. We found that the contrast of the 10 nm CoO film is about 2% at 290 K, and reaches about 9.3% at 77 K, but the birefringence contrast of a 20 nm NiO film is only ~2.5% at room temperature [13], so the CoO film contains much stronger optical birefringence effect than the NiO film grown on MgO(001). Such difference may be attributed to the different spin orientation in NiO and CoO film, since the CoO AFM spins lie in the film plane [19,29], and the NiO AFM spins align along the canting directions with an angle of ~10° away from the normal direction [13].

Fig. 4(b) shows the temperature dependent contrast for the CoO films with the two thicknesses of 2.3 nm and 6.6 nm, which clearly decreases with the temperature and vanishes at high temperature. The temperature dependence of the contrast can be well fitted by the characteristic $\langle M \rangle_T^2$ behavior of mean filed theory, which results in the $T_N$ of 268 K for 2.3 nm thick CoO, and ~330 K for 6.6 nm thick CoO. Our results demonstrate that the magneto-optical birefringence effect can be used to determine the ordering temperature of AFM thin films.

We further determine $T_N$ as a function of CoO film thickness, shown in Fig. 4(c). The thickness dependent $T_N$ is expected to follow the finite-size scaling relation as [18,33,34]

$$\frac{T_N(\infty) - T_N(d)}{T_N(\infty)} = \left(\frac{\xi_0}{d}\right)^\lambda \quad (1).$$

Here, $T_N(\infty)$ is the Néel temperature in the bulk, $T_N(d)$ is the $T_N$ of the film with a finite thickness $d$, $\xi_0$ is the extrapolated spin-spin correlation length at zero temperature, and $\lambda$ is the shift exponent for the finite-size scaling [18,33]. Our data can be well fitted with Eq. (1), resulting in the fitting parameters of $\lambda = 2.4 \pm 0.1$, $\xi_0 = (1.24 \pm 0.06)$ nm and $T_N(\infty) = (335 \pm 14)$ K. Fig. 4(d) shows the log-log plot of $(T_N(\infty) - T_N(d))/T_N(\infty)$ as a function of CoO thickness, and further proves the power law dependence of $T_N$ in Eq. (1). The finite size effect of the CoO ordering temperature has been proved by the specific heat measurements in polycrystalline CoO films [34], magnetic susceptibility in CoO/SiO$_2$ superlattices[18], inverse spin Hall effect in YIG/CoO/Pt heterostructures [35], and the exchange coupling in Fe/CoO bilayers [19]. Our results prove that the magneto-optical birefringence effect can be used to identify the AFM finite size effect from a wedge film, which will potentially benefit further study on the AFM thin films. We note that the fitted $T_N(\infty)$ for CoO in our study is higher than the reported values ranging between 293 K and 315 K in polycrystalline CoO thin film [34-36] or in CoO/oxide superlattices [18,33], which is likely induced by the epitaxial strain from the MgO substrate, since theoretical calculations indicated that the lattice strain can enhance the $T_N$ of AFM Cr$_2$O$_3$ [37,38].

We also find that the optical birefringence effect of CoO film strongly depends on the wavelength of incident light. Figures 5(a)-(c) show the same AFM domains of a 4.6 nm thick CoO measured by the three LED sources with different colors. We found that the domain contrast is the strongest for the blue LED, but the weakest for the red LED. The polarization rotation angle $\theta_v$ can be quantified by measuring the asymmetry $I_{asy}$ as a function of the analyzer's offset angle $\theta_P$ due to the relation of $I_{asy} = 2\theta_v/\theta_P$ [13]. Figure 5(d) shows the $\theta_P$-dependent $I_{asy}$ measured with three LED sources, which is inversely proportional to $\theta_P$ [Fig. 5(e)]. The determined $\theta_v$ of the 4.6 nm CoO film is 168.5 mdeg for blue LED, 91.0 mdeg for white LED and 38.5 mdeg for red LED, so the optical birefringence effect of CoO film increases with the photon energy [Fig. 5(f)]. The measured $\theta_v$ of the 4.6 nm CoO is much larger than that from NiO films, which is 60 mdeg from a 20 nm NiO film measured with the white LED [13]. It should be noted that $\theta_v$ of CoO film is much larger than the longitudinal Kerr angles from Fe [39] or Co [40] thick films, which are usually less than 21 mdeg.

The CoO(001) film has the in-plane four-fold symmetry, thus it is expected that two AFM domains with the spins along <110> and <1$\bar{1}$0> have the similar fractional areas. However, we found one domain type dominates in our CoO films, indicating the spins along <110> and <1$\bar{1}$0> axis are not energetically equilibrium. This observation can be attributed to the atomic steps of MgO(001) substrate, and the substrate miscut is very difficult to be avoided. X-ray reflectivity measurement reveals a small miscut angle of ~0.5 degree with the step along <110> axis in the sample in Fig. 2, which can induce the uniaxial anisotropy in CoO film with the easy axis along the <110> direction [24], as well as the dominated white CoO domains in Fig. 2.

It is worthy to further discuss the origin of the AFM domains in CoO film. It is known that domain walls in both FM and AFM films will increase the exchange coupling energy and anisotropy energy. However, due to the zero net magnetization in AFM systems, the formation of AFM domain will not reduce the magnetostatic energy, so the intrinsic AFM domain is not energetically favorable in a perfect AFM crystal system [41]. Thus, the observed CoO AFM domains are likely induced by the local strains or defects in the CoO film, which can explain the fact that the domain structures in pure CoO region in Fig. 2 always keep the same after different field cooling processes. Nevertheless, our results show the possibility to create a single AFM domain in CoO film through exchange coupling, thus it is also possible to generate the AFM domain wall in the CoO nanostructures and understand the motion of AFM domain walls induced by the magnetic field or electronic current, which is the desired information for the AFM spintronics application.

## IV. Conclusion

In summary, we demonstrate the AFM domains in single crystalline CoO ultrathin films grown on MgO(001) can be observed by magneto-optical birefringence effect. The single AFM domain in Fe/CoO(001) bilayer can be achieved through the exchange coupling, and the spin orientation in CoO AFM domains can be manipulated by the external cooling field. By analyzing the domain contrast, we further prove that the thickness-dependent Néel temperature of ultrathin CoO film follows the finite size effect. The magneto-optical birefringence effect of CoO film is found to increase with the photon energy of incident light, and a large polarization rotation angle of 168.5 mdeg is determined from a 4.6 nm CoO film with a blue light source. Since AFM domains in both CoO and NiO [13] films have been successfully imaged, the magneto-optical birefringence effect can be considered as a general method to study the AFM domain distribution in different AFM materials. Since the optical imaging method can adapt with external magnetic fields or electric currents, the AFM domains can be imaged during the operation of electric current or magnetic field, thus the magneto-optical birefringence effect can be applied to understand the mechanism of AFM domain switching [2-7] or AFM domain wall motion [42,43] induced by current or field pulses, which is helpful for future developments of AFM spintronics devices.

**Acknowledgements**

This work was supported by the National Key Research and Development Program of China (Grant No. 2016YFA0300703), National Natural Science Foundation of China (Grant No. 11974079 and 11734006), and the Program of Shanghai Academic Research Leader (No. 17XD1400400). G. C. acknowledges support by the NSF (DMR-1610060) and the UC Office of the President Multicampus Research Programs and Initiatives (MRP-17-454963).

**Figures:**

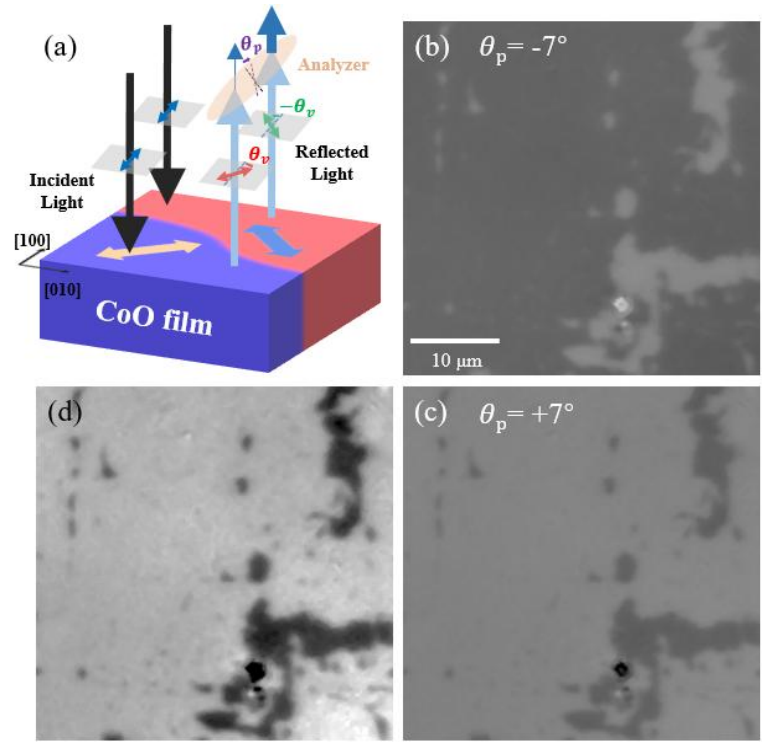

Fig.1. (a) Schematics of magneto-optical microscopy measurement geometry. (b)-(c) The optical image of a 5.5 nm CoO obtained at (b) $\theta_P = -7°$ and (c) $\theta_P = +7°$ respectively. $\theta_P$ is the rotation angle of the analyzer from the extinction position, and the positive (negative) sign of $\theta_P$ means the clockwise (anti-clockwise) rotation. (d) The contrast image calculated by the signal asymmetry, i.e. $I_{asy} = \frac{I(+\theta_P) - I(-\theta_P)}{I(+\theta_P) + I(-\theta_P)}$.

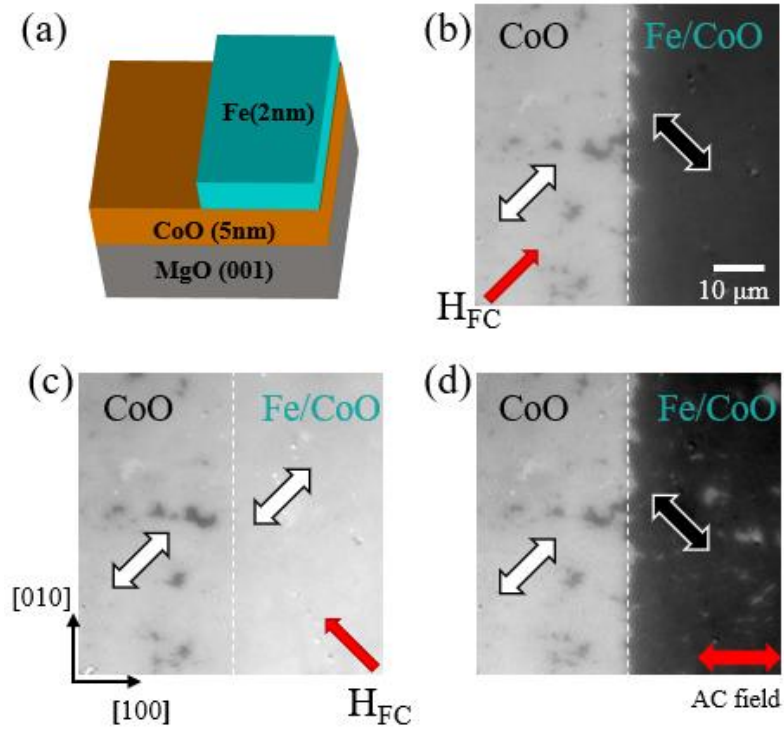

Fig. 2. (a) The sample structure with a pure 5 nm CoO layer and a Fe(2 nm)/CoO(5 nm) bilayer. (b)-(c) The birefringence images at 77 K after field cooling along (b) [110] and (c) [$\bar{1}$10]. The red arrows show the directions of the cooling field $H_{FC}$. (d) The birefringence image by zero field cooling after the film is demagnetized at RT by an AC field along <100>. The dash white line in (b)-(d) is the guide line to show the boundary between pure CoO and Fe/CoO regions. The black and white arrows show the spin directions of CoO.

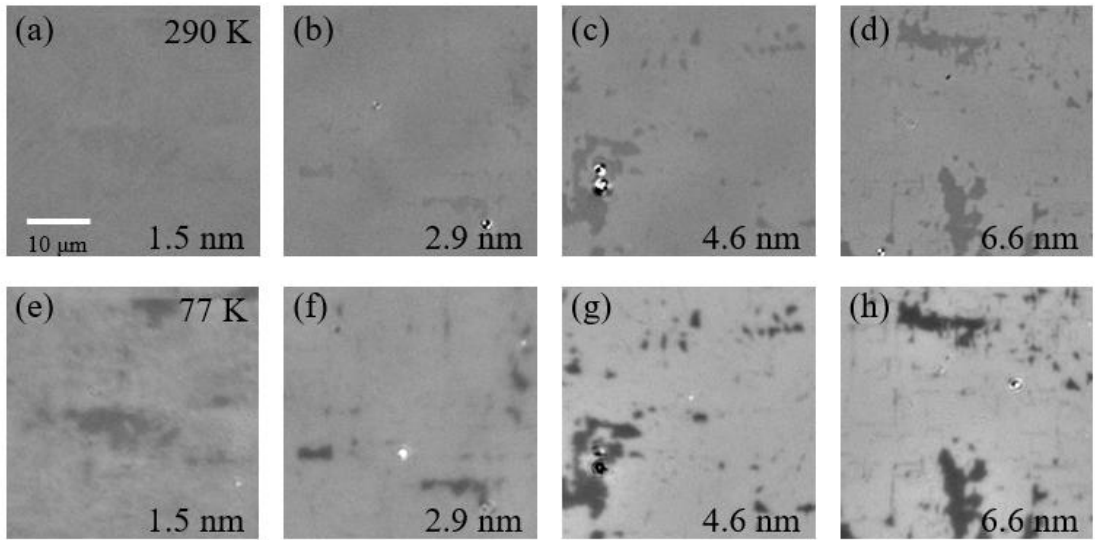

Fig. 3. Thickness-dependent measurements of AFM domains in pure CoO films at (a)-(d) 290 K and (e)-(h) 77 K.

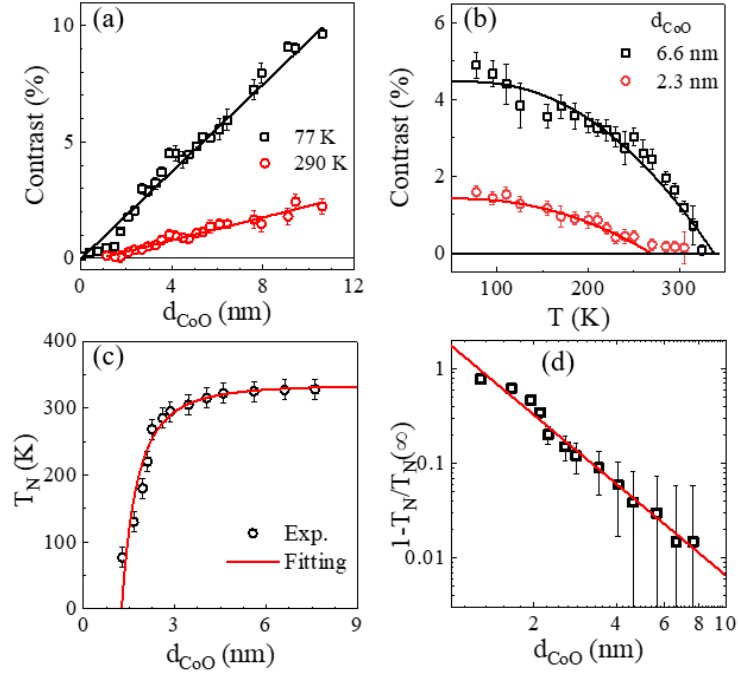

Fig. 4. The birefringence contrast as a function of (a) CoO thickness at different temperatures and (b) temperature for different thicknesses. The solid lines in (a) are the straight lines guide to eyes, and the solid lines in (b) are the theoretical temperature dependence as discussed in the text. (c) Thickness-dependent $T_N$ of pure CoO films. (d) Log-log plot of $[1-T_N/T_N(\infty)]$ vs CoO thickness. The red lines in (c) and (d) are the fitting curves with Eq. 1.

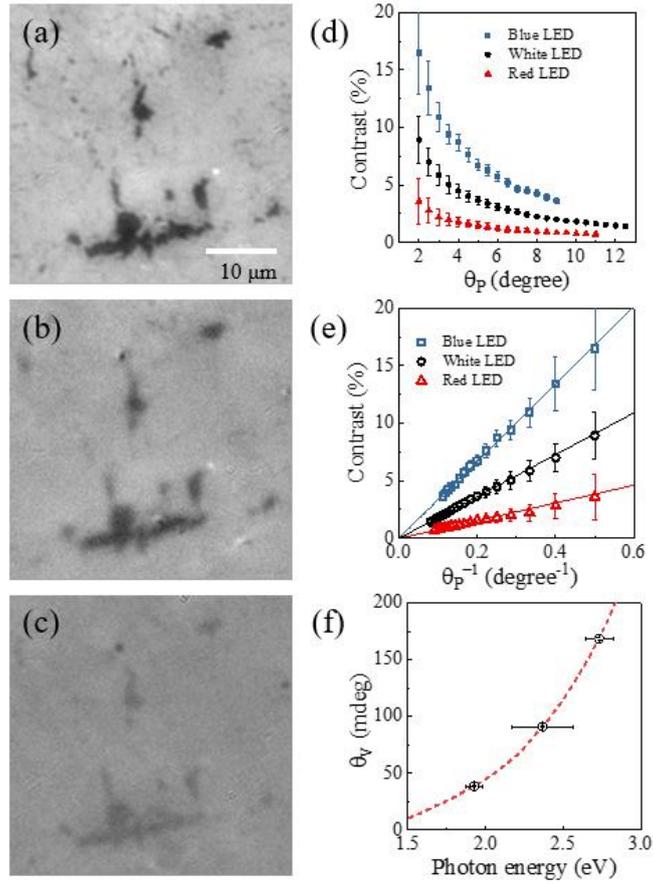

Fig. 5. (a)-(c) The optical birefringence images from a 4.6 nm CoO layer measured at 77 K with different LED sources: (a) blue LED, (b) white LED and (c) red LED. (d)-(e) The birefringence asymmetry $I_{asy}$ as a function of (d) $\theta_P$ and (e) $\theta_P^{-1}$ for different LED sources. (f) The determined polarization rotation angle $\theta_v$ as a function of photon energy. The dashed line is a guide to eyes.